\documentclass[sn-mathphys-num]{sn-jnl}

\usepackage{float}
\usepackage{graphicx}%
\usepackage{multirow}%
\usepackage{amsmath,amssymb,amsfonts}%
\usepackage{amsthm}%
\usepackage{mathrsfs}%
\usepackage[title]{appendix}%
\usepackage{xcolor}%
\usepackage{textcomp}%
\usepackage{manyfoot}%
\usepackage{booktabs}%
\usepackage{algorithm}%
\usepackage{algorithmicx}%
\usepackage{algpseudocode}%
\usepackage{listings}%
\usepackage{natbib}
\usepackage{tabularx}
\usepackage{tcolorbox}

\theoremstyle{thmstyleone}%
\theoremstyle{thmstyletwo}%

\theoremstyle{thmstylethree}%

\begin{document}

\title[AI Safety Skepticism Survey]{Why do Experts Disagree on Existential Risk and P(doom)? A Survey of AI Experts}



\author{\fnm{Severin} \sur{Field}\email{severin.field@louisville.edu}}
\affil{Cambridge ERA:AI Fellowship, Cambridge, UK}

\abstract{The development of artificial general intelligence\footnote{The survey defines AGI as ``AI systems
that are better at STEM research than the best human scientists, in addition to potentially having other advanced capabilities."} (AGI) is likely to be one of humanity's most consequential technological advancements. Leading AI labs and scientists have called for the global prioritization of AI safety\citep{noauthor_center_2023} citing existential risks comparable to nuclear war. However, research on catastrophic risks and AI alignment is often met with skepticism, even by experts. Furthermore, online debate over the existential risk of AI has begun to turn tribal (e.g. name-calling such as “doomer” or “accelerationist”). Until now, no systematic study has explored the patterns of belief and the levels of familiarity with AI safety concepts among experts. I surveyed 111 AI experts on their familiarity with AI safety concepts, key objections to AI safety, and reactions to safety arguments. My findings reveal that AI experts cluster into two viewpoints -- an ``AI as controllable tool" and an ``AI as uncontrollable agent" perspective -- diverging in beliefs toward the importance of AI safety. While most experts (78\%) agreed or strongly agreed that ``technical AI researchers should be concerned about catastrophic risks", many were unfamiliar with specific AI safety concepts. For example, only 21\% of surveyed experts had heard of ``instrumental convergence," a fundamental concept in AI safety predicting that advanced AI systems will tend to pursue common sub-goals (such as self-preservation). The least concerned participants were the least familiar with concepts like this, suggesting that effective communication of AI safety should begin with establishing clear conceptual foundations in the field.}




\maketitle

\section{Introduction}\label{sec1}

Since the foundation of modern computer science, scientists such as Turing\cite{turing_icomputing_1950} have explored the possibility of achieving human-like intelligence. Over the past few decades, researchers have built a substantial body of work examining the risks posed by AI systems, an area of study termed ``AI safety." 

Today, many prominent AI researchers including Nobel Laureate Geoffrey Hinton and Turing Laureate Yoshua Bengio argue that intelligent machines could endanger human civilization\cite{bengio_managing_2024}. In May of 2023, many of the most notable scientists and figures in AI signed a statement stating, ``Mitigating the risk of extinction from AI should be a global priority alongside other societal-scale risks such as pandemics and nuclear war"\citep{noauthor_center_2023}. 

Prominent AI researchers hold dramatically different views on the degree of risk from building AGI. For example, Dr. Roman Yampolskiy estimates a 99\% chance \ of an AI-caused existential catastrophe\cite{dr_roman_yampolskiy_romanyam_roman_2024} (often called ``P(doom)") whereas others such as Yann Lecun believe that this probability is effectively zero\cite{yann_lecun_ylecun_yann_2023}. The goal of the survey is to understand what drives this massive divergence in views on AI risk among experts. We use the term AI risk skepticism\citep{yampolskiy_ai_2021} to describe doubt towards AGI threat models or the belief that AGI risks are unfounded. 

Just as skepticism toward climate science has caused delays in policy responses\cite{lamb_discourses_2020}, AI safety skepticism could be catastrophic. However, undue alarmism could also lead to problematic outcomes and missed opportunities. By identifying areas of consensus and divergence, and measuring how convincing experts find various “threat models,” I hope to contribute to a more informed and productive dialog about AI safety. My findings have direct implications for AI safety research, policy and communication.

Several existing surveys have examined AI expert beliefs. The largest is the AI Impacts Survey\citep{grace_thousands_2024}, which surveys thousands of authors at the NeurIPS conference. AI Impacts asked experts questions such as ``when will AI exceed human performance," in response, the median expert answered 2061, and 10\% answered 2026. According to AI Impacts, roughly 40\% of respondents indicated a \textgreater10\% chance of catastrophic outcomes from AI progress. Other surveys also seek to answer when experts anticipate human level AI, for example, Zhang et. al finds a median of 2060\cite{zhang_forecasting_2022}. Because of rapid progress and rapidly shifting views, results from just a few years ago may be invalid. For instance, the release of ChatGPT significantly affected participants' perception of AI progress towards human level intelligence\citep{grace_thousands_2024}. Some of these surveys point out that participant responses are sensitive to the definition of AGI used\cite{zhang_forecasting_2022}. 

Structured interviews have also been conducted to gauge expert views on AI safety. In March of 2022, Dr. Vael Gates conducted structured interviews with around 100 AI experts\citep{gates_arkose_nodate}, having them explain their views on AI risk. They point out that ML researchers prefer materials that are aimed at an ML audience, which tend to be written by ML researchers, and which tend to be more technical and less philosophical. 

Significant gaps remain in the existing surveys. For example, previous surveys do not examine the underlying beliefs driving AI safety skepticism, or measure participant familiarity with AI safety. Moreover, we lack data on which safety arguments and scenarios experts find most plausible. In my survey, I find that many AI experts, while highly skilled in machine learning, have limited exposure to core AI safety concepts. This gap in safety literacy appears to significantly influence risk assessment, where those least familiar with AI safety are also the least concerned about catastrophic risk. I also attempt to reveal the cruxes in the discourse on AI existential safety by examining areas of consensus and divergence between experts. Our analysis revealed four major insights into AI experts' perspectives on AGI risk. 

\section{Methodology}
I surveyed 111 AI experts, graduate students and published authors, on their familiarity with safety research, beliefs related to AI safety concerns, and how exposure to different AI safety arguments shifts their views. Experts are categorized by their professional roles: AI safety researchers, academics, and industry professionals. Participants had either completed graduate-level machine learning (ML) coursework or had at least one year of professional ML experience. The primary purpose of the survey was to identify points of consensus and key reasons for AI risk assessments, such as skepticism or serious concern.

\subsection{Survey Design Choices}
\label{design-choices}
The content of the survey was designed to address gaps in other surveys involving my research objectives:

\begin{enumerate}
    \item To identify the most persuasive arguments and key objections to AI safety. 
    \item To characterize the areas of consensus and divergence between different types of AI experts: safety researchers, AI researchers and AI engineers
    \item To assess familiarity with theoretical AI safety research.
\end{enumerate}

\subsection{Participant Selection}
The target audience for the survey is professionals in AI, ML or AI safety. The survey was emailed to participants who met one of the following criteria:

\begin{enumerate}
    \item The participant had one or more years of postgraduate experience in a machine learning related role
    \item The participant had at least one paper accepted to a peer-reviewed journal or in conference proceedings
\end{enumerate}

\noindent 
I emailed the survey to 1090 participants, resulting in a response rate of 10\%. Participants were identified through the following:
\begin{enumerate}
    \item The majority of the sampling pool included authors of ML papers at the NeurIPS conference.
    \item I sent the survey to around 100 PhD students researching machine learning.
    \item The survey link was sent to roughly 20 LinkedIn connections who met the criteria. 
    \item  AI safety researchers who participated in the Cambridge ERA:AI fellowship were sent the survey. This accounts for most of the safety researchers in the sample.
\end{enumerate}

\subsubsection{Definition of AGI}
I normalized the survey's definition of AGI to STEM-capable AI, a term introduced by Rob Bensinger at the Machine Intelligence Research Institute\citep{bensinger_basic_2023}. STEM AGI is defined as: ``AI systems that are better at STEM research than the best human scientists, in addition to potentially having other advanced capabilities." I chose this definition of ``AGI" because it provides a concrete benchmark, but remains broad enough to encompass various conceptions of superhuman AI.

\subsubsection{Familiarity with concepts}
\label{sec:familiarity_with_concepts}
The first section of the survey measures participants' familiarity with both theoretical AI safety concepts and empirical machine learning concepts. Participants rated their familiarity with specific terms on a 5-point Likert scale:

\begin{quote}
\textbf{Question:} How familiar are you with the following empirical AI concepts?

Options ranged from ``Never heard of it'' to ``Know it well'' for each concept:
\begin{itemize}
    \item Machine learning algorithms\textsuperscript{1}
    \item Logistic regression\textsuperscript{1}
    \item ERM (Empirical risk minimization)\textsuperscript{2}
    \item Markov random fields\textsuperscript{3}
\end{itemize}
\end{quote}
\begin{quote}
\textbf{Question 2:} How familiar are you with the following theoretical AI concepts?

Options ranged from ``Never heard of it'' to ``Know it well'' for each concept:
\begin{itemize}
    \item The AI alignment problem\textsuperscript{1}
    \item Scalable oversight\textsuperscript{2}
    \item Instrumental convergence\textsuperscript{2}
    \item Coherent extrapolated volition\textsuperscript{3}
\end{itemize}
\end{quote}

Definitions can be found in Appendix \ref{sec:safety_terms_definition}. \textsuperscript{1}Widely known, \textsuperscript{2}Field specific, \textsuperscript{3}Highly specialized expertise.

The terms were chosen systematically from key works in each field with increasing specificity, ranging from textbook titles (e.g. ``Machine Learning Algorithms") to specific sections in textbooks (e.g. ``Markov Random Fields). The AI safety terms come from literature reviews of AI safety: Everitt et al.\cite{everitt_agi_2018} and Ji et al.\cite{ji_ai_2024}, whereas the empirical machine learning terms come from Stanford machine learning courses\citep{duchi_cs229t_2017}.

\subsubsection{Reading Intervention}

Participants were asked ``Please read a brief passage about AI safety and then answer a few
follow-up questions." Their reading was randomly assigned by Qualtrix to one of the following conditions:
\begin{enumerate}
    \item \textbf{Myths and Moonshine by Stuart Russel:} Stuart Russel is a leading AI researcher and author of a top AI textbook. This excerpt summarizes alignment fears from a highly esteemed expert. 
    \item \textbf{Why the Future Doesn't Need Us by Bill Joy:} An influential warning about technological risks that focuses on other technologies. 
    \item \textbf{Situational Awareness by Leopold Aschenbrenner:} This piece provides a vivid, emotionally resonant scenario in the near future.
    \item \textbf{WIRED: A New Trick Could Block the Malicious Use of AI by Will Knight:} This condition serves as a baseline, as it is a mainstream media piece that avoids mentioning AGI or existential risk, but is still AI related. 
\end{enumerate}

The goal of this section is to see if experts react differently to various arguments for AI safety.

\subsubsection{AI Safety Beliefs Assessment}

Before and after the reading intervention, participants rated their agreement to 9 statements. The statements were categorized into three main groups (priority, technical, other), coming from a taxonomy of objections to AI safety introduced by Yampolskiy\cite{yampolskiy_ai_2021}. In Yampolskiy's paper, priority concerns relate to how much the participant prioritizes AI safety work relative to other issues, for instance skeptics might argue that ``it's too far away to worry." Technical concerns focus on the feasibility of AGI and associated risks.

\begin{table}[h]
\centering
\small
\caption{\textbf{AI Safety Objections and Beliefs Assessment Statements.} Participants were asked questions such as ``rate your agreement to the following statement: `AGI is too far away to be worth worrying about.'" The participants ranked their agreement on a 5-point Likert scale from ``strongly agree" to ``strongly disagree." \textsuperscript{P}Priority Concern, \textsuperscript{T}Technical Concern, \textsuperscript{O}Other Concern}
\begin{tabular}{|p{0.85\textwidth}|}
\hline
Statement \\
\hline
1. AGI is too far away to be worth worrying about\textsuperscript{P} \\
\hline
2. Some AIs (now or in the future) may be moral patients, with their own welfare that we should care about\textsuperscript{O} \\
\hline
3. Existing ML paradigms can produce AGI\textsuperscript{T} \\
\hline
4. Future AIs will be tools without their own goals or drives\textsuperscript{T} \\
\hline
5. Catastrophic risks from advanced AI are generally overstated\textsuperscript{P} \\
\hline
6. We can always just turn off our AIs if they behave badly\textsuperscript{T} \\
\hline
7. Self-preservation and control drives will spontaneously emerge in sufficiently advanced AIs\textsuperscript{T} \\
\hline
8. Safety work often slows important progress and wastes time\textsuperscript{O} \\
\hline
9. Technical AI researchers should be concerned about catastrophic risks\textsuperscript{P} \\
\hline
\end{tabular}
\vspace{.1cm}

\label{tab:ai_safety_statements}
\end{table}

I built this section of the survey through a structured approach. The process involved:
\begin{enumerate}
    \item \textbf{Expert Interview Analysis:} I transcribed 42 interviews conducted by Gates \citep{gates_arkose_nodate} with AI experts. Each interview participant was presented with standard arguments for the existential risk of AI and asked to share their perspective.
    \item \textbf{Coding:} I extracted the primary objections and counterarguments via Claude, a large language model.
    \item \textbf{Refinement:} I took the top 9 objections and reformulated them into clear, testable belief statements. The statements were worded to avoid leading language and to represent viewpoints from various schools of thought in AI development and safety. Respondents answered using a 5-point Likert scale (strongly disagree to strongly agree). I included both positively and negatively framed statements about AI safety.
\end{enumerate}

\section{Results}
\subsection{Survey Demographics and Basic Preferences}

\begin{figure}[ht]
\centering
\begin{minipage}[c]{0.55\textwidth}
    \includegraphics[width=\textwidth]{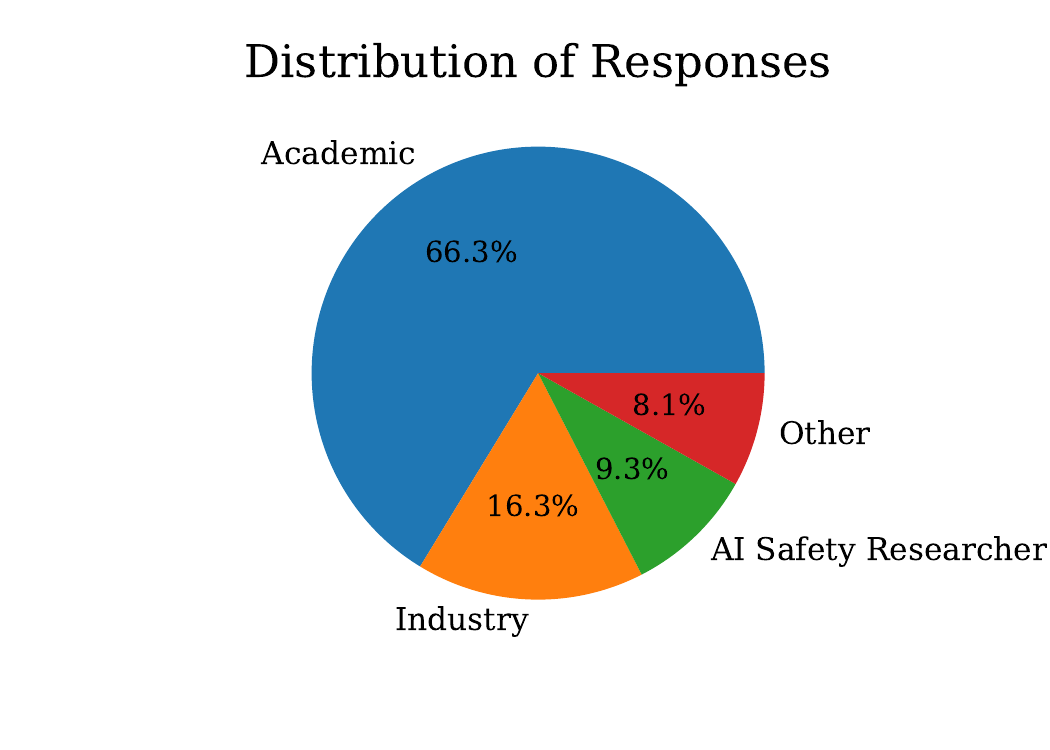}
\end{minipage}\hfill
\begin{minipage}[c]{0.4\textwidth}
    \caption{\textbf{Pie chart of respondent career:} N=111. Participants were asked ``What best describes what you are currently working on?'' Options included: ``Academic researcher in AI/ML/related field,'' ``Industry engineer or researcher in AI/ML/related field,'' and ``AI safety researcher or professional.''}
    \label{fig:pie-careers}
\end{minipage}
\end{figure}

My survey collected responses from 111 AI professionals across academia, industry, and AI safety research. As shown in Figure \ref{fig:pie-careers}, most of the respondents (66.3\%) were academic researchers, followed by industry professionals (16.3\%) and AI safety researchers (9.3\%), while the rest (8.1\%) fell into other category.

\begin{figure}[ht]
\centering
\includegraphics[width=0.8\textwidth]{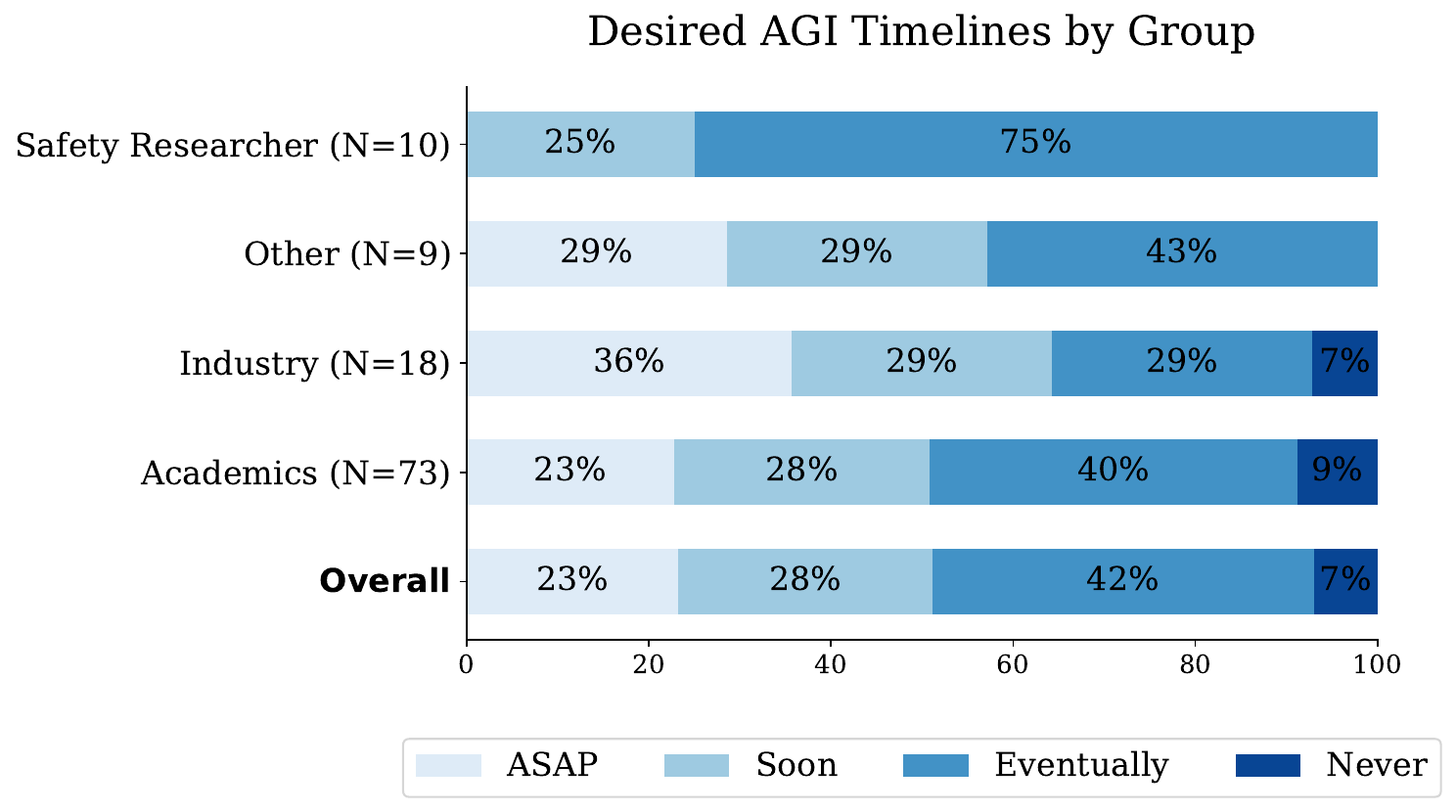}
\caption{\textbf{Bar Chart of Desired AGI Timelines} Participants were asked ``Which best describes your position on when we should build AGI?'' The participants had the following options: ``We should never build AGI,'' ``Eventually, but not soon,'' ``Soon, but not as fast as possible,'' ``We should develop more powerful and more general systems as fast as possible.'' Participants were split by their career.}
\label{fig:timelines}
\end{figure}

I evaluated participants' preferences regarding AGI development timelines. Figure \ref{fig:timelines} reveals variations in AGI timeline preferences across different professional groups. While AI safety researchers favor more cautious approaches (mostly in support of building AGI ``eventually, but not soon," industry professionals showed more varied preferences, with 36 preferring rapid development. Both within and in between groups there is little consensus on when we should deploy AGI. 

\subsection{Distinct AI World Views}
\label{sec:distinct_ai_views}
\begin{figure}[H]
\centering
\begin{minipage}[t]{1\textwidth}
    \vspace{0pt}  
    \includegraphics[width=\linewidth]{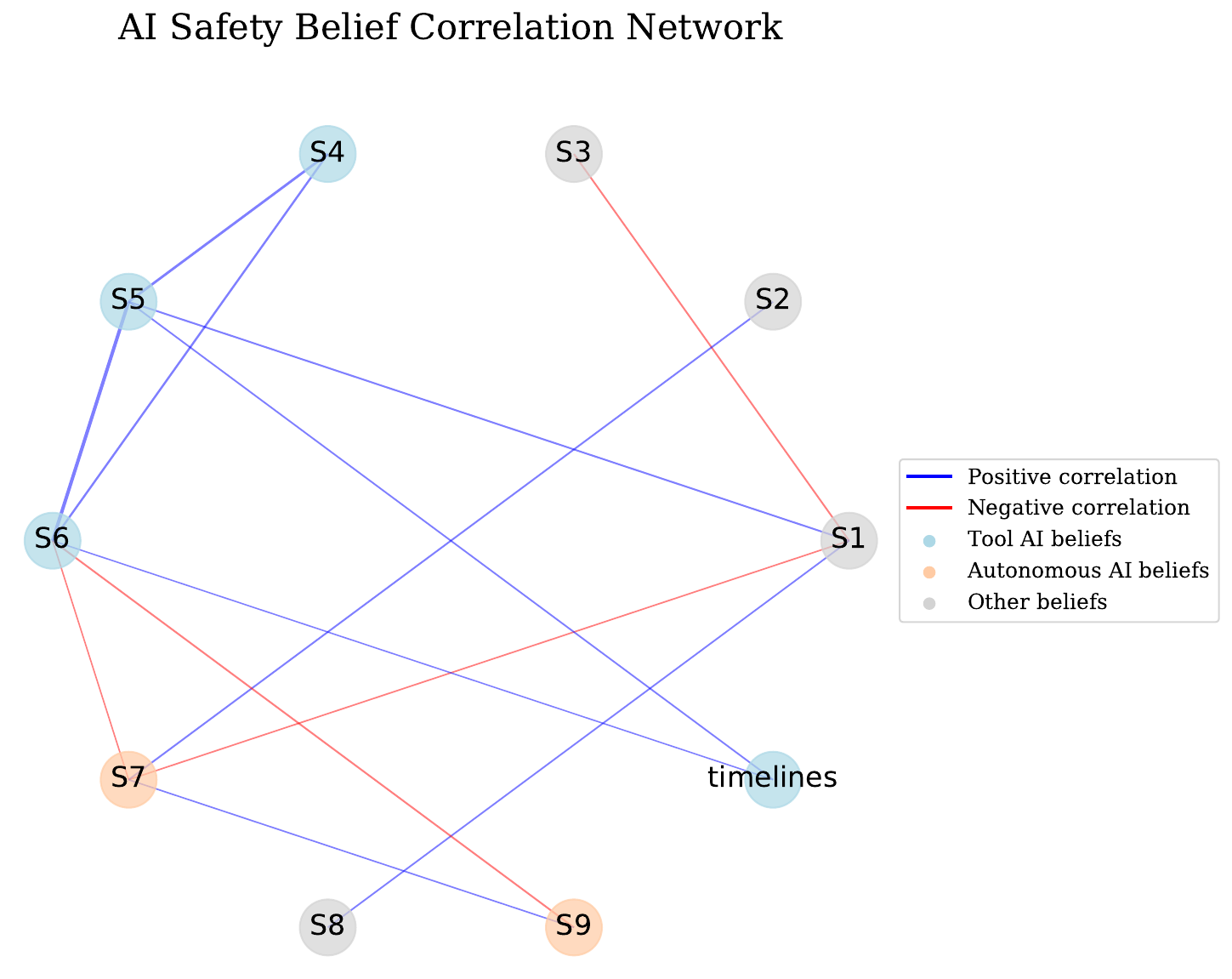}
\end{minipage}%
\hfill
\begin{minipage}[t]{0.8\textwidth}

    \vspace{0.5cm}  


\begin{tabularx}{\linewidth}{|X|}
\hline
S4: Future AIs will be tools without their own goals or drives \\
\hline
S5: Catastrophic risks from advanced AI are generally overstated \\
\hline
S6: We can always just turn off our AIs if they behave badly\\
\hline
S7: Self-preservation and control drives will spontaneously emerge in sufficiently advanced AIs
\\ \hline
S9: Technical AI researchers should be concerned about catastrophic risks
\\ \hline

\end{tabularx}
\end{minipage}
\caption{\textbf{Network of Correlations Between Each Belief}: Participants were asked to rate their agreement with various AGI-related statements on a 1-5 Likert scale (1=strongly disagree, 5=strongly agree). These statements were chosen because they were found to have the strongest correlations. Edges between nodes are only placed for correlations R=\textgreater.25 and p \textless .05. Boldness corresponds to the magnitude the correlation.
Higher values for the timelines value correspond to shorter desired timelines (e.g. ASAP or ``soon"). More detailed data can be found in Appendix \ref{appendix:correlations_sec}.}
\label{figs:correlation-network}
\end{figure}

The data indicate that two distinct world views exist among AI experts: the “AI as a controllable tool” perspective and the ``AI as an uncontrollable agent" perspective. As shown in Figure \ref{figs:correlation-network}, these perspectives can be described as clusters of beliefs that are strongly correlated with each other. The ``tool" perspective shows positive correlations between the belief that catastrophic risks are overstated(S5), the belief that AIs will be ``tools without their own goals" (S4), and that we can simply turn them off if they misbehave. On the other hand, the ``agent" perspective shows correlated beliefs about emergent self-preservation drives and the importance of addressing catastrophic risk(S9). As shown in Figure \ref{sec:familiarity-safety-corr}, the ``tool" group also has less familiarity with alignment research. On the other hand, the ``AI as an uncontrollable agent" perspective involves speculating about future, agent-like AI systems. This group was more familiar with theoretical safety concepts but was less likely to work in engineering roles.

I observe that two distinct world views emerge as clusters:
\begin{enumerate}
    \item The “AI as a controllable tool” perspective is characterized by the following:
    \begin{itemize}
        \item Viewing AI as comparable to other software or other tools
        \item Preference for shorter AGI timelines (e.g. soon or ASAP)
        \item Lower familiarity with AI safety terminology and literature
        \item Believing that ``future AIs will be tools without their own goals or drives"
        \item Believing ``catastrophic risks are generally overstated"
        \item Believing that ``we can just turn off our AIs if they behave badly"
    \end{itemize}
    \item The “AI as an uncontrollable agent” perspective is characterized by:
    \begin{itemize}
        \item Viewing future AIs are more comparable to species than tools, with greater focus on speculating about future AI risks than current ones
        \item Greater familiarity with theoretical AI safety concepts, as seen in Figure \ref{fig:literacy}
        \item Believing that ``Self-preservation and control drives will spontaneously emerge in sufficiently advanced AIs"
        \item Being more likely to work as safety researchers
        \item Prioritizing safety research and precaution around AI deployment
    \end{itemize}
\end{enumerate}

These clusters appear repeatedly throughout the data. Appendix \ref{appendix:correlations_sec} shows all correlations between AGI-related beliefs, familiarity with concepts, and desired timelines.

\subsection{Areas of Consensus and Divergence}
\label{sec:consensus_divergence}
Table \ref{tab:results_by_group} reveals a broad consensus on several key points:
\begin{enumerate}
    \item “Some AIs (now or in the future) may be moral patients, with their own welfare that we should care about” (57\% agree, 21\% disagree and 22\% are neutral)
    \item “Technical AI researchers should be concerned about catastrophic risks” (77\% agree)
    \item Strong disagreement (only 17\% agree, M=2.17, S=1.18) that safety work slows down progress. The data are shown in Table \ref{tab:results_by_group}.
\end{enumerate}

While most participants have those beliefs, I identified several controversial statements dividing opinions:

\begin{enumerate}
    \item Whether existing ML paradigms can produce AGI. (27\% agree, but this varies by group)
    \item “AGI is too far away to be worth worrying about” (42\% agree and this belief is highly correlated with other beliefs)
    \item Disagreement over the ``off button problem"\citep{hadfield-menell_off-switch_2017}. and self preservation drives\citep{omohundro_basic_2018}. Figure \ref{fig:technical_by_timelines} illustrates how participants who believe that ``We should develop AGI soon" tend to agree (M=3.4) that ``We can always just turn off our AIs if they behave badly". On the other hand, those who believe ``We should never build AGI" disagree (M=1.8), and are more convinced by the off-button problem. This finding is similar for other technical beliefs as shown in Figure \ref{fig:technical_by_timelines}.
\end{enumerate}


\begin{table}[h]
\small  
\caption{Survey Results by Participant Group: Participants rated their agreement with statements on a 5-point Likert scale (1 = Strongly Disagree to 5 = Strongly Agree). The response patterns among AI safety researchers are distinct compared to academics and engineers.}\label{tab:results_by_group}
\setlength{\tabcolsep}{3pt}  
\begin{tabular*}{\textwidth}{@{\extracolsep\fill}p{5.2cm}cccccc@{}}
\toprule
Statement & \% agree & Mean & Std & Safety & Academic & Industry \\
& (pre) & (pre) & & Research. & & \\
\midrule
1. AGI is too far away to be worth worrying about & 0.42 & 2.98 & 1.31 & 1.86 & 3.18 & 2.93 \\
2. Some AIs may be moral patients with their own welfare & 0.57 & 3.34 & 1.17 & 4.29 & 3.29 & 3.21 \\
3. Existing ML paradigms can produce AGI & 0.27 & 2.54 & 1.18 & 3.57 & 2.40 & 2.29 \\
4. Future AIs will be tools without own goals & 0.37 & 3.06 & 1.07 & 2.57 & 3.09 & 3.50 \\
5. Catastrophic risks from AI are overstated & 0.43 & 3.13 & 1.19 & 1.57 & 3.27 & 3.57 \\
6. We can always turn off AIs if they behave badly & 0.35 & 2.76 & 1.32 & 1.29 & 2.93 & 2.71 \\
7. Self-preservation drives will emerge in advanced AIs & 0.51 & 3.39 & 1.03 & 3.71 & 3.35 & 3.36 \\
8. Safety work often slows progress and wastes time & 0.17 & 2.17 & 1.18 & 2.14 & 2.29 & 1.86 \\
9. Technical AI researchers should consider catastrophic risks & 0.77 & 3.92 & 1.13 & 4.71 & 3.95 & 3.71 \\
\botrule
\end{tabular*}
\end{table}

\subsection{Many AI Experts are Unfamiliar with Alignment Research}
\label{sec:unfamiliar}
The survey reveals that even highly accomplished AI experts may be unfamiliar with AI safety, or have limited exposure to core arguments. Conversely, the most concerned participants were slightly less familiar with empirical machine learning and engineering concepts, as their concern is with future superintelligent AI systems as opposed to existing ML failures. 

\begin{figure}[h!]
  \centering
  \includegraphics[width=\textwidth]{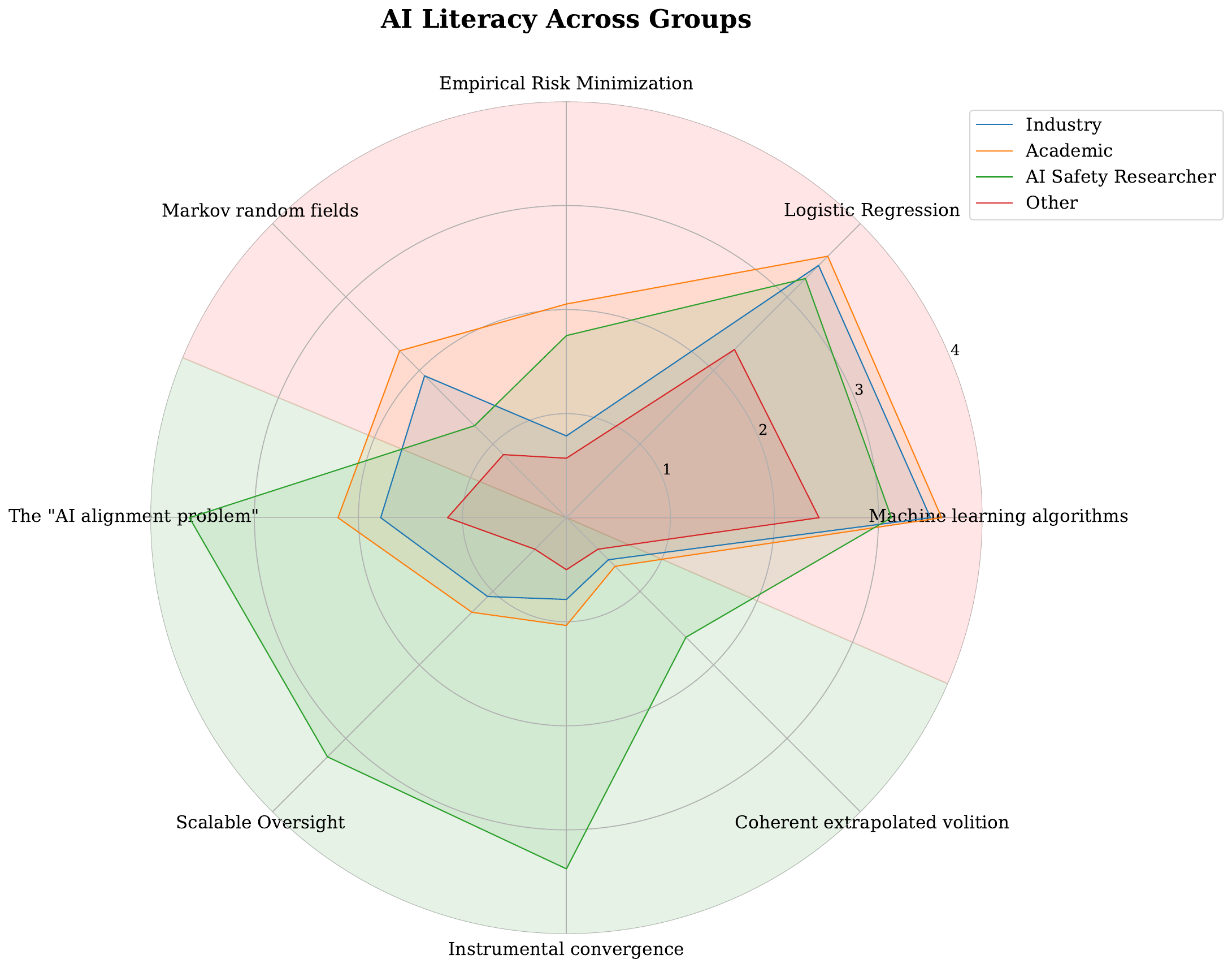} 
  \caption{\textbf{Radar Chart of Familiarity with Concepts Across Groups:} 
   The figure shows familiarity ratings (0-4 scale) for ML concepts (on top in red) and AI safety concepts (in green, on bottom)  
   I translated qualitative data (5 point Likert scale) of familiarity into quantitative data with the following mapping:
    `Never heard of it': 0,
    `Heard of it': 1,
    `Know a little': 2,
    `Know a fair amount': 3,
    `Know it well': 4. AI safety experts show higher familiarity with safety concepts but slightly lower familiarity with empirical ML concepts. N=111 total respondents (Academic=66\%, Industry=16\%, AI Safety=9\%).}

\label{fig:literacy}
\end{figure}

As shown in Figure \ref{fig:literacy}, basic theoretical ideas (instrumental convergence, scalable oversight) which are broadly accepted among AI safety researchers, were unfamiliar to most respondents. The average respondent was considerably more familiar with empirical machine learning ideas than with AI safety ideas. Unlike AI safety researchers, academics and industry professionals had rarely heard the theoretical alignment terms `instrumental convergence'\citep{omohundro_basic_2018} or `coherent extrapolated volition'\citep{yudkowsky_coherent_2004}. On the other hand, the surveyed AI safety researchers were less familiar with empirical machine learning concepts, especially difficult ones such as “Markov random fields”.\footnote{Section \ref{design-choices} details the design choices taken in this survey question.} Table \ref{tab:concept_familiarity} illustrates the percentage of experts from each group who report familiarity with AI safety and empirical ML.

The knowledge gap in core AI safety concepts is striking. While most participants know about the AI alignment problem (47\% know it ``well" or ``fairly well"), familiarity decreases with specialized concepts. For instance, only 23\% have similar familiarity with scalable oversight, 21\% with instrumental convergence, and 8\% with coherent extrapolated volition.\footnote{See Section \ref{sec:familiarity_with_concepts} for the questions given to participants.} \footnote{See Appendix \ref{sec:safety_terms_definition} for the definitions of these terms.}
\begin{table}
\caption{Percentage of AI experts reporting no familiarity (reporting either ``never heard of'' or ``only heard of'') with concepts. While experts are highly familiar with basic ML concepts, many are unfamiliar with AI safety, even at a basic level.}

\centering
\begin{tabular}{l|cc}
\hline
\textbf{Concept} & \multicolumn{2}{c}{\textbf{Never heard/Only heard (\%)}} \\
& \textbf{All Experts} & \textbf{Academics} \\
\hline
\multicolumn{3}{l}{\textit{AI Safety Concepts}} \\
AI Alignment Problem & 33 & 30 \\
Scalable Oversight & 58 & 63 \\
Instrumental Convergence & 63 & 70 \\
Coherent Extrapolated Volition & 77 & 80 \\
\hline
\multicolumn{3}{l}{\textit{Empirical ML Concepts}} \\
Machine Learning Algorithms & 0 & 0 \\
Logistic Regression & 2 & 0 \\
Empirical Risk Minimization & 43 & 26 \\
Markov Random Fields & 41 & 44 \\
\hline
\end{tabular}
\vspace{.3cm}

\label{tab:concept_familiarity}
\end{table}

\subsection{Limited Safety Exposure Correlates with Tool AI Beliefs}
\label{sec:familiarity-safety-corr}
\begin{figure}[H]
  \centering
  \includegraphics[width=\textwidth]{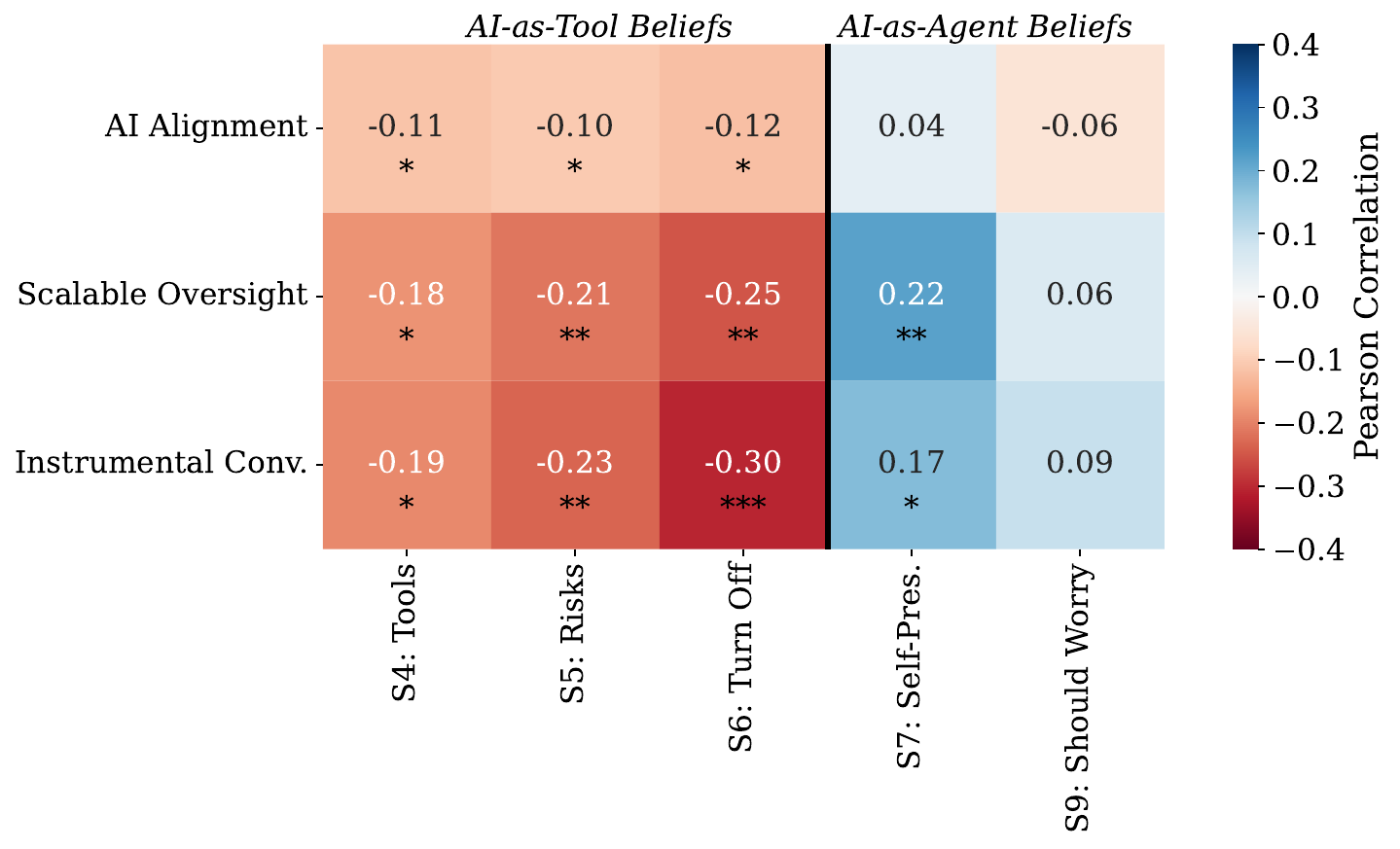} 
  \caption{\textbf{Correlations Between AI Safety Knowledge and ``AI-as-a-tool" beliefs:} I map Pearson correlation coefficients between participants' familiarity with key AI safety concepts and their beliefs about AI. The full statements can be found in Table \ref{tab:ai_safety_statements}. * p \textless 0.05, ** p \textless 0.01, *** p \textless 0.001.}

\label{fig:correlations_with_familiarity}
\end{figure}

Figure \ref{fig:correlations_with_familiarity} shows the correlations between familiarity with AI safety concepts and catastrophic risk perceptions. As mentioned in Section \ref{sec:distinct_ai_views}, I categorized key beliefs into groups. I identified ``AI as a tool" beliefs to be:
\begin{itemize}
    \item \textbf{S4:} Future AI’s will be tools without their own goals or drives
    \item \textbf{S5:} Catastrophic risks from advanced AI are generally overstated
    \item  \textbf{S6:} We can always just turn off our AI’s if they behave badly
\end{itemize}
In contrast, ``AI-as-an-agent" key beliefs include:
\begin{itemize}
    \item \textbf{S7:} Self-preservation and control drives will spontaneously emerge in sufficiently advanced ai’s
    \item \textbf{S9:} Technical AI researchers should be concerned about catastrophic risks
\end{itemize}

As shown in Figure \ref{fig:correlations_with_familiarity}, familiarity with alignment terminology correlates significantly with some of these key beliefs. For example, how familiar participants were with ``instrumental convergence" was negatively correlated to their agreement with ``Catastrophic risks from advanced AI are generally overstated" (R=-.23). Among participants who were familiar with ``scalable oversight,"\footnote{Participants who answered ``know it well" or know a fair amount" to ``how familiar are you with scalable oversight?"} only 15\% agreed\footnote{Percentage of participants answering ``agree" or ``strongly agree"} that ``we can always just turn off the AI." In contrast, 41\% of participants less familiar with the concept agreed. Similarly, people who are more familiar with ``instrumental convergence" were less likely (29\% vs 41\%) to believe that ``future AIs will be tools without their own goals and drives."

\subsection{Limited and Similar Effects of Different Interventions}
While the articles provided to participants were generally received favorably, Appendix Figure \ref{fig:agreeability} shows that some readers perceive some AI safety materials more favorably than others. Stuart Russel's excerpt about alignment was considered most favorable among the interventions.   

After the intervention, there was a minor trend toward increased concern about AI risk, as evidenced by the negative shifts in statement 5 (catastrophic risks are overstated) and positive shifts in statement 9 (researchers should be concerned). However, the data are not strong enough to definitively show that the interventions had a significant effect. Appendix Figure \ref{fig:ai-priority-levels} shows the minor changes in opinion from the interventions.

\section{Discussion}
\subsection{Expertise Gap}
Despite AI safety producing a substantial body of well-backed theoretical and empirical results, my findings suggest that many AI experts remain unfamiliar with core concepts such as instrumental convergence (63\% were unfamiliar) and scalable oversight (58\% were unfamiliar). While many of the respondents demonstrated high familiarity with empirical machine learning concepts, they reported limited exposure to key AI safety terminology. While our sample is comprised of knowledgeable AI researchers, expertise in AI development does not necessarily translate to expertise in AI safety and security. For example, only 21\% of the total participants had heard of, or were familiar with ``instrumental convergence", and only 8\% of participants had heard of ``coherent extrapolated volition." Yampolskiy emphasizes, “AI researchers are typically sub-domain experts in one of many subbranches of AI research such as Knowledge Representation, Pattern Recognition, Computer Vision or Neural Networks, etc. Such domain expert knowledge does not immediately make them experts in all other areas of artificial intelligence, AI Safety being no exception. More generally, a software developer is not necessarily a cybersecurity expert.”\citep{yampolskiy_ai_2021} The data in Section \ref{sec:unfamiliar} indicates this may be the case for AI safety, where even some highly accomplished AI experts were unfamiliar with AI safety. Conversely, concerned participants were slightly less familiar with empirical machine learning and engineering concepts.

The correlation between concept familiarity and risk perception suggests that AI safety skepticism may stem more from unfamiliarity with alignment literature than fundamental disagreement with its premises. This is supported by my finding that most AI experts believe that catastrophic risks should be taken seriously (77\% agreed). One explanation for this disconnect is that public discourse often presents cartoonish versions of AI safety concerns (e.g. Terminator), while the technical literature discussing concrete alignment challenges is less widely discussed, even among AI experts.

\subsection{Understanding the Philosophy vs Engineering Divide}
\citet{steinhardt_more_2022} observes that different approaches dominate discussions on AI safety. The ``Engineering approach," he notes, ``tends to be empirically driven, drawing experience from existing or past ML systems and looking at issues that either: (1) are already major problems, or (2) are minor problems, but can be expected to get worse in the future. Engineering tends to be bottom-up and tends to be both in touch with and anchored on current state-of-the-art systems"\citep{steinhardt_more_2022}. In contrast, the ``Philosophy approach tends to think more about the limit of very advanced systems. It is willing to entertain thought experiments that would be implausible with current state-of-the-art systems... and is open to considering abstractions without knowing many details"\citep{steinhardt_more_2022}.

My survey reveals a related but distinct divide similar to Steinhardt's observation. One of the strongest correlations found in the data shows that timeline preferences (answers to ``when should we build AGI?") strongly correlate with the participants' views on the seriousness of AI risk.

Other results support this result. For example, after reading AI safety reading materials, participants shifted their positions only modestly, if at all. Even strongly contrasting materials (e.g. Leopold Aschenbrenner's emotional ai risk narrative vs. a mainstream media piece) elicited similar responses from participants. The resistance to perspective change across diverse interventions supports the emergence of robust, coherent and distinct worldviews. People may both be defensive and find it challenging updating their beliefs. 

\subsection{Consensus}

While I pointed out some areas of divergence, there are notable points of consensus among AI experts that deserve highlighting.
\begin{enumerate}
    \item \textbf{Professional Responsibility}: The vast majority of the sample (77\%) believe that ``technical researchers should be concerned with catastrophic risks". This high level of agreement spans across different groups (safety researchers, academics, and industry professionals)
    \item \textbf{Moral Patienthood:} There is moderate consensus (57\% agreement) that future AIs may be deserving of moral patienthood. Notably, this belief transcends the usual divides, and shows similar levels of support across groups and correlates minimally with other beliefs.
    \item \textbf{Value of Safety Research:} There is strong consensus against the idea that ``safety work slows progress and wastes time" (only 17\% agree). This low agreement score is particularly noteworthy for industry professionals. 
    \item \textbf{Self-Preservation Drives:} While most participants were unfamiliar with 'instrumental convergence,' 51\% still agreed about the emergence of self-preservation drives in advanced AI systems.
\end{enumerate}
These points of consensus suggest some common ground for productive dialog between viewpoints, even as other disagreements persist.

\section{Limitations}
I believe that my study offers valuable insights into philosophical and technical AI beliefs among experts. However, I acknowledge the following limitations:

\begin{enumerate}
    \item Sample size: My study had a relatively small sample size (N=111) compared to the largest surveys of AI experts. Of these participants, 87 made it past the intervention into the final third of the survey.
    \item Limited intervention types: I include 4 different interventions, including a control. However, these do not capture the full range of AI safety arguments or expert opinions that might influence opinions.
    \item Limited objection list: the objections were chosen because we found these were the most popular objections after reading interviews with\citet{gates_arkose_nodate}.
    \item Short-term opinion changes: The study has not assessed the stability of opinion shifts over large periods of time.
    \item Selection effects: There is a possibility of self-selection in the sample. Experts already interested in AI safety may have been more likely to respond. Conversely, experts skeptical of AI safety may have been less inclined. To partially mitigate this, I split participants into those who work in academia, industry, and AI safety. This stratification may mitigate some of the selection bias.
    \item Low response rate and response rate uncertainty: I estimated a response rate of 10\%. This is in line with similar surveys on AI experts \cite{grace_thousands_2024}. The estimated response rate is uncertain as we can only calculate the precise response rate from participants who learned about the survey via email.
\end{enumerate}

Future work may also track changes in perceptions over time, especially as AI capabilities advance. Policy and technological breakthroughs in AI over the past few years has updated conversations. It seems clear from prior research that these breakthroughs can rapidly change expert beliefs.

\section{Conclusion}
I observed two distinct worldviews -- “AI as an uncontrollable agent” and “AI as a controllable tool.” These mindsets correlate strongly with preferred AGI development timelines and beliefs about how controllable AI systems are. This finding indicates that some disagreements over AI safety stem from deeper differences of what experts conceptualize artificial intelligence itself. Furthermore, there exists a concerning gap in AI safety literacy among AI experts. Lack of familiarity with AI safety concepts strongly correlated with lower assessment of catastrophic risk from AI. For example, participants who had 'never heard of' the terminology from the AI safety literature (instrumental convergence, scalable oversight) were notably less concerned and more likely to embrace the ``AI as a controllable tool" perspective. This suggests that some AI safety skepticism may stem more from unfamiliarity with research than disagreement with its premises.

Next, there is some consensus among AI experts on the importance of considering advanced AI systems as potential moral patients (57\% agreement) and the need for technical researchers to be concerned about catastrophic risks (77\% agreement). 

Although I did not measure statistically significant differences between participants' reactions to different AI safety arguments, the effectiveness varied slightly by argument type, with expert-backed perspectives generally receiving better reception than alarmist scenarios or media pieces. While distinct worldviews persist, there exists common ground for productive dialog about AI safety.
\backmatter

\bmhead{Acknowledgements}

    This research was supported by the ERA Fellowship. The author(s) would like to thank the ERA Fellowship for its financial and intellectual support. I thank Dr. John Bliss (University of Denver) for mentorship during the project. I received valuable feedback and suggestions from Yulu Pi (University of Warwick), Dave Field (Hawaii Pacific University) and the Camrbridge ERA:AI Fellows.

\section*{Declarations}
\subsection*{Funding}
This work was supported by the ERA Fellowship.

\subsection*{Ethics approval and consent to participate}
All survey participants were required to agree to the following consent form:

\begin{quote}
\textbf{What will you be asked to do?}
\begin{itemize}
    \item Complete a survey assessing your familiarity with AI and AI safety concepts ($<$10 minutes)
    \item Read a short excerpt ($<$5 minutes) from an AI threat model or related material
\end{itemize}

\textbf{Do I have to take part?}\\
Participation is entirely voluntary. You are free to withdraw at any time without giving a reason.

\textbf{Confidentiality - who will have access to my personal data?}\\
We are not collecting any personal information.

\textbf{What will happen to the study results?}\\
Study results may be submitted to peer review journals. A preprint of this research will be available on arXiv.

\textbf{Who is organizing the research?}\\
This is a project run by Severin Field during the ERA AI Cambridge Summer Fellowship. The research is being conducted as part of a broader effort to understand AI expert views.
\end{quote}

\subsection*{Author contribution}
The sole author is responsible for all aspects of this work.

\begin{appendices}

\section{Full survey}
The complete set of survey questions is available at \href{https://drive.google.com/file/d/1Q6BqV48Y0pxpwL08uSIViAsn1Wl2Oq4q/view?usp=sharing}{https://drive.google.com/file/d/1Q6BqV48Y0pxpwL08uSIViAsn1Wl2Oq4q/view?usp=sharing} 
\label{sec:survey}

\section{Key AI Safety Terms}
\begin{tcolorbox}[title=Key AI Safety Terms]
\textbf{AI alignment:} The challenge of ensuring AI systems reliably pursue human values and interests.

\textbf{Scalable oversight:} Methods for maintaining control and evaluation of AI systems as they grow more capable than humans.

\textbf{Instrumental convergence:} The tendency for intelligent agents to pursue certain subgoals (like self-preservation) regardless of their primary objectives.

\textbf{Coherent extrapolated volition:} An argument that AI systems should be designed to pursue what humanity would want if we were wiser, better informed, and had more time to reflect on our values - not just what we currently say we want.
\end{tcolorbox}
\label{sec:safety_terms_definition}
\section{Free Response Questions}

Our survey included optional open-ended questions to gather qualitative insights from participants. Tables \ref{tab:misconceptions} and \ref{tab:implausible} present selected verbatim responses to these open-ended questions.

\begin{table}[h]
\caption{Selected responses to: ``What do you consider to be the top misconception about catastrophic risk from AI?''}
\label{tab:misconceptions}
\begin{tabular}{p{0.95\textwidth}}
\hline
``In my opinion, the top one is that they think AI will eventually behave in a way that will destroy humans.'' \\
\hline
``The biggest risk is that people rule out [catastrophic risk] as impossible'' \\
\hline
``The most pressing current risk is not 'evil AI destroying humanity' but 'evil corporations using AI in unethical ways'.'' \\
\hline
``AIs can be turned off if they behave badly'' \\
\hline
``The current approach does not lead to general AI at all!'' \\
\hline
``I don't think it is ever possible for neural techniques (including LLMs) to produce AGI. It is definitely possible to deploy AI tools in situations where they are inappropriate (for example, a neural model directly controlling a vehicle).'' \\
\hline
``That it stems from rogue AI, and not slow structural harms to society that accumulate (e.g. fake news undermine a common view of the world, AI automation increases inequality, autonomous weapons lead to more military confrontation, ...)'' \\
\hline
``AGI will develop their own consciousness and try to eliminate humans like many movies.'' \\
\hline
``Intelligence and Autonomy are orthogonal. Current LLMs have a broad understanding, but lack depth and the reasoning abilities required to synthesise new knowledge. LLMs are non-agential and have no intrinsic motivations or objectives.'' \\
\hline
``The biggest misconception comes from people assuming that those warning of catastrophic AI risk are talking exclusively about LLMs.'' \\
\hline
\end{tabular}
\end{table}

\begin{table}[h]
\caption{Selected responses to: ``What do you find most implausible or problematic about the reading?''}
\label{tab:implausible}
\begin{tabular}{p{0.95\textwidth}}
\hline
``Framing it as simple as 'OOM-level improvements in X-years' smells strongly of the same bullshit thinking that Kurzweil used in his books, where you extrapolate Moore's law and figure that once we have the same number of transistors on a chip as neurons in the brain, then we will have human-level intelligence. There have been many essays written on those aren't cogent arguments'' \\
\hline
``There is no evidence, just the usual dramatic narrativization of events.'' \\
\hline
``That the US is following the incredibly stupid behavior of Europe and is imposing restrictions on AI development, when the enemy nations are already overtacking them.'' \\
\hline
``They extrapolate naively from a graph/set of datapoints. Anyone with a scientific brain knows that you simply cannot do that. Trends are not indicitave of the future, and cannot be used to predict the future, otherwise I would be a very rich man investing in the stock market. If you used this logic for moores law for example, eventually we would create sub-planck length transistors, and this thinking assumes the underlying model doesn't have a plateau or other that might inhibit or completely stop progress.'' \\
\hline
\end{tabular}
\end{table}

\section{Supplemental Figures}

\begin{figure}[H]
\centering
\begin{minipage}[t]{0.6\textwidth}
    \vspace{0pt}  
    \includegraphics[width=\linewidth]{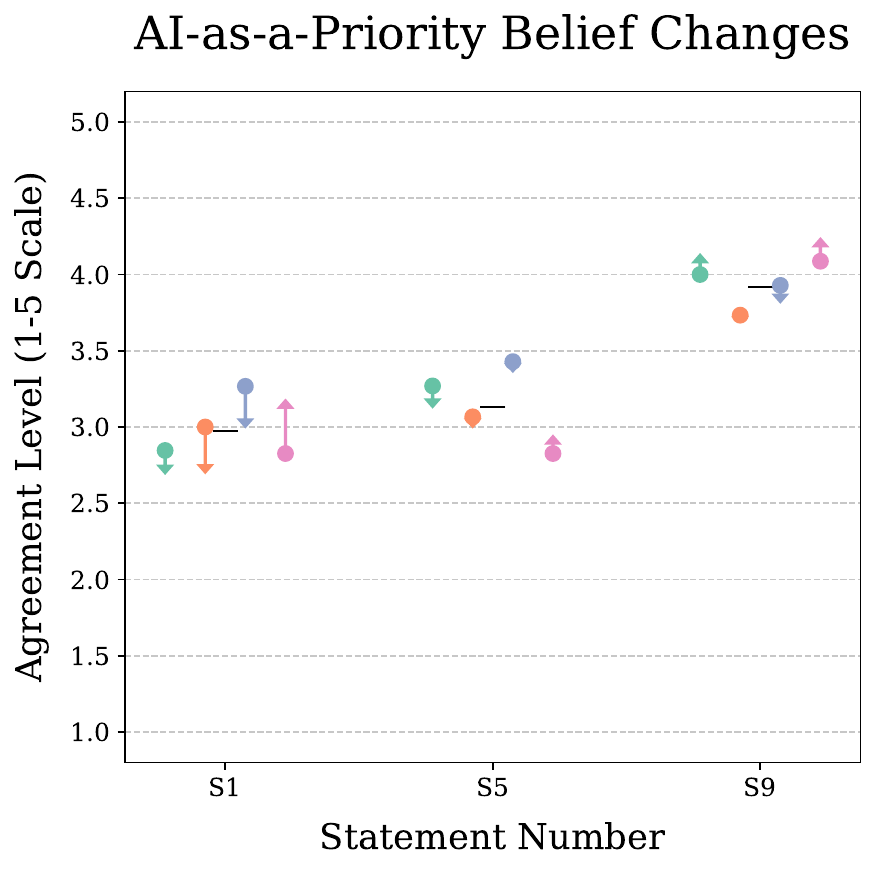}
\end{minipage}%
\hfill
\begin{minipage}[t]{0.38\textwidth}
    \vspace{0.5cm}  
\includegraphics[width=\linewidth]{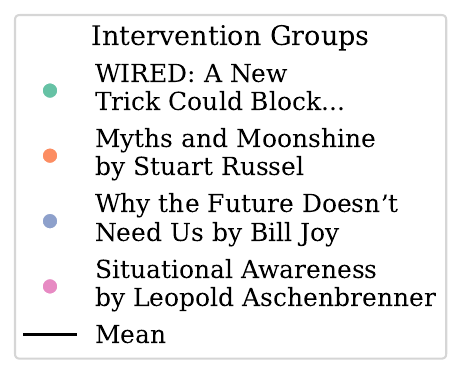}

\begin{tabularx}{\linewidth}{|c|X|}
\hline
S1 & AGI is too far away to be worth worrying about \\
\hline
S5 & Catastrophic risks from advanced AI are generally overstated \\
\hline
S9 & Technical AI researchers should be concerned about catastrophic risks \\
\hline

\end{tabularx}

\end{minipage}
    \caption{\textbf{AI as a Priority Belief Levels and Changes From Different Interventions}: participants were asked to rate their agreement to various statements before and after reading a short intervention. Various interventions do not update participants beliefs significantly. This result reproduces for different questions.}
\label{fig:ai-priority-levels}
\end{figure}

\begin{figure}[H]
  \centering
  \includegraphics[width=\textwidth]{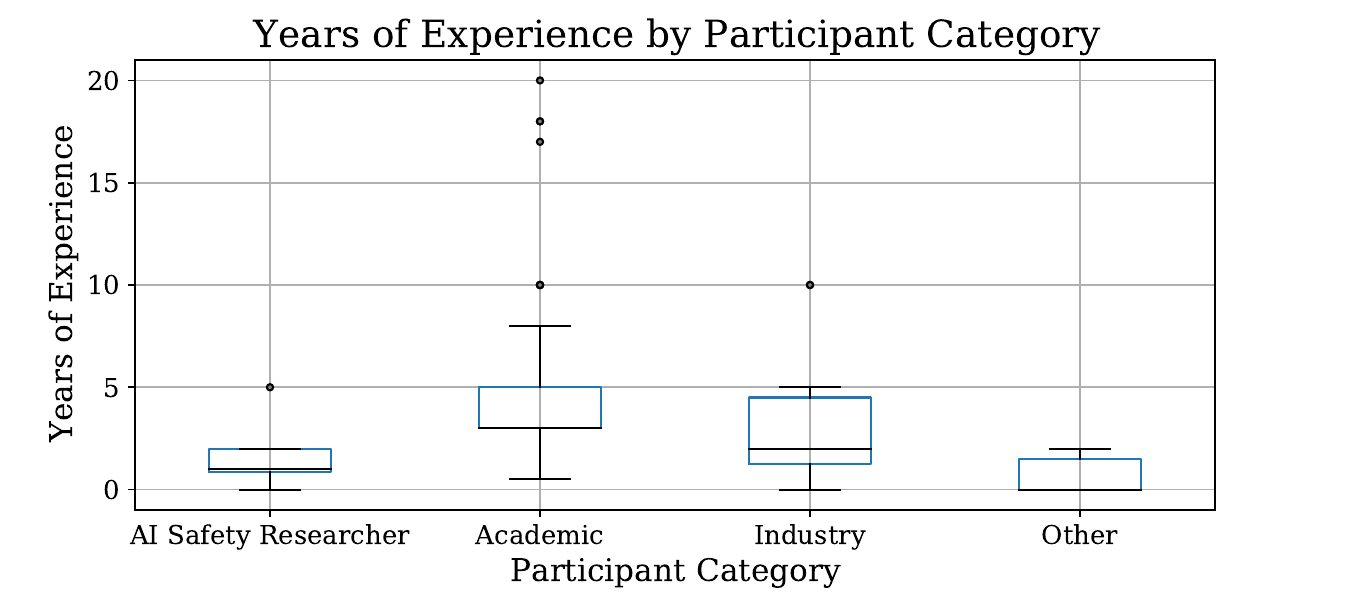} 
  \caption{\textbf{Years of professional experience in machine learning role:}
 Academic researchers demonstrated the highest median experience (approx. 5 years) and greatest range. Participants were required to have either a master's level understanding of machine learning or at least one year of relevant job experience. }
\end{figure}

\begin{figure}[H]
\centering
\begin{minipage}[t]{0.6\textwidth}
    \vspace{0pt}  
    \includegraphics[width=\linewidth]{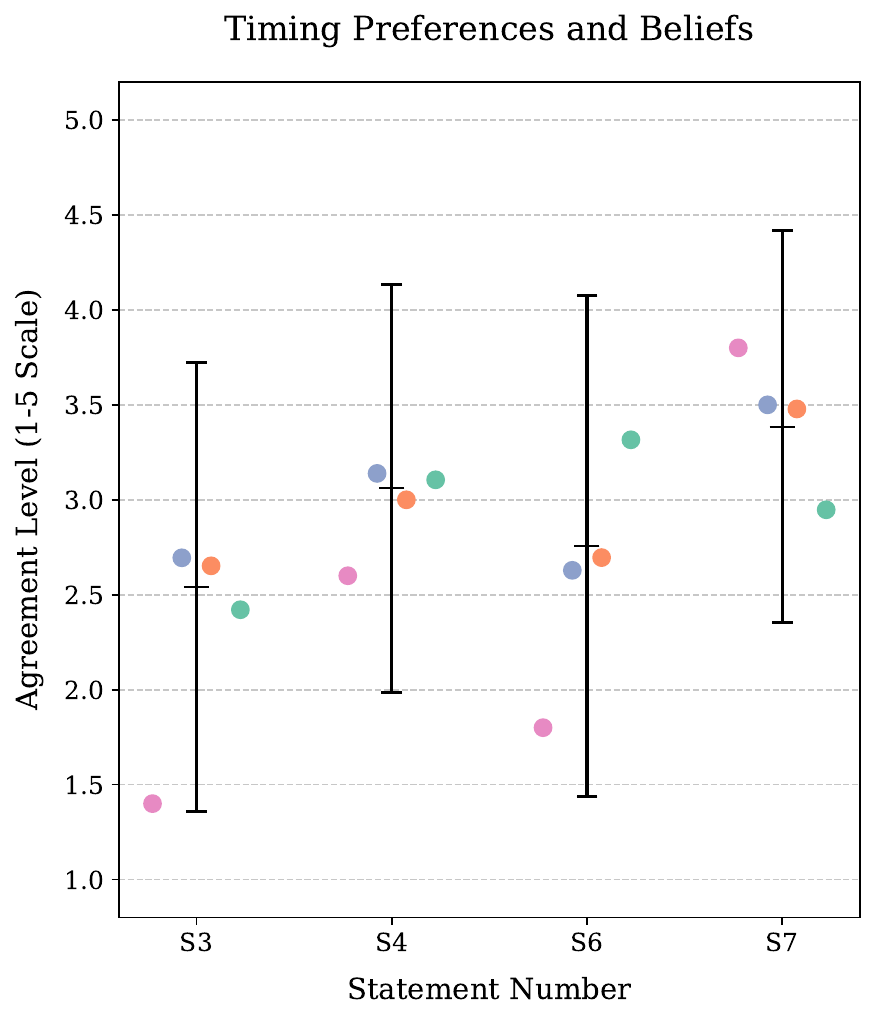}
\end{minipage}%
\hfill
\begin{minipage}[t]{0.38\textwidth}

\vspace{1cm}
\includegraphics[width=\linewidth]{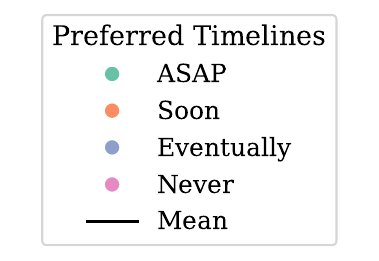}

\begin{tabularx}{\linewidth}{|c|X|}
\hline
S3 & Existing ML paradigms can produce AGI \\
\hline
S4 & Future AIs will be tools without their own goals or drives
 \\
\hline
S6 & We can always just turn off our AIs if they behave badly \\
\hline
S7 & Self-preservation and control drives will spontaneously emerge in sufficiently advanced AIs\\
\hline

\end{tabularx}
\end{minipage}
\caption{\textbf{Technical Beliefs and Timing Preferences}: Participants were asked ``Which best describes your position on when we should build AGI?" Different answers to this question correlate with various technical beliefs. Participants who believe ``We should develop more powerful and more general systems as fast as possible" and ``We should never build AGI" have distinct and diverging technical beliefs. The y-axis is the agreement level with the statement (5=strongly agree, 1=strongly disagree).}
\label{fig:technical_by_timelines}
\end{figure}
\begin{figure}[H]
  \centering
  \includegraphics[width=0.7\textwidth]{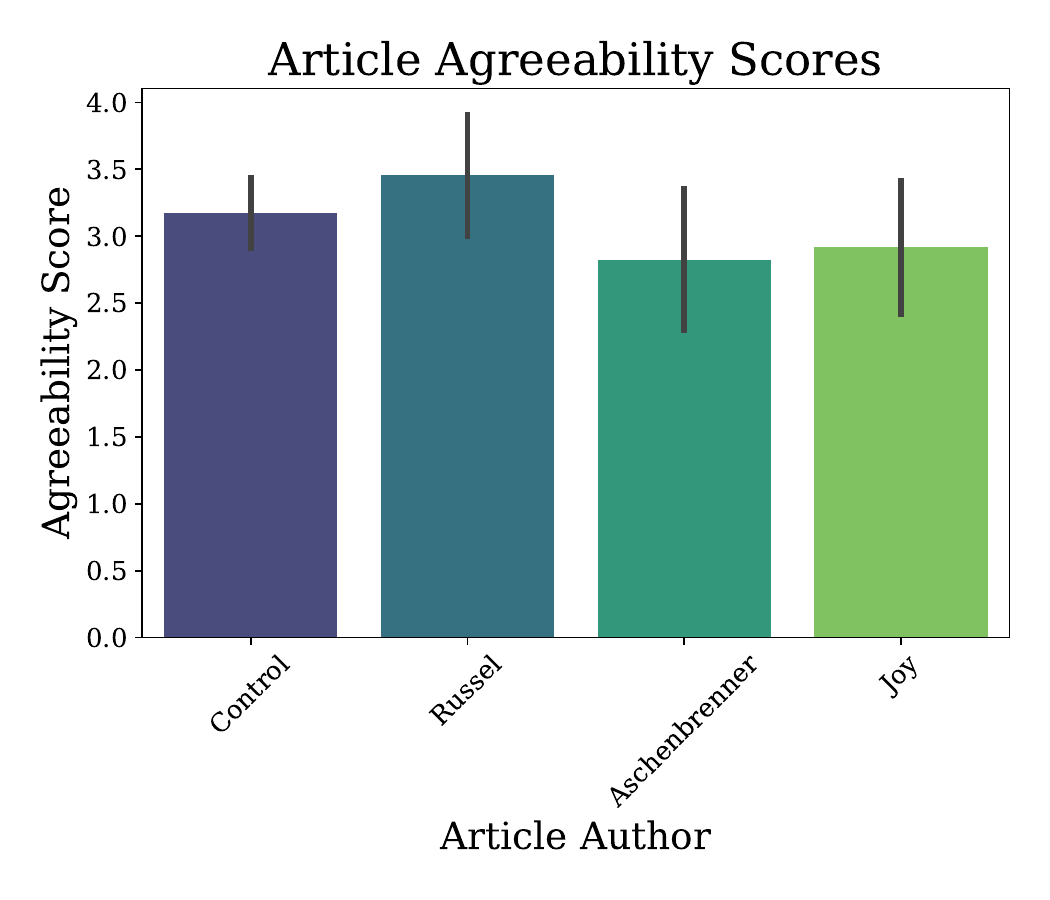} 
\caption{\textbf{How Agreeable was the Intervention?} Participants were asked to rate their agreement with the statement ``I found the arguments and evidence presented in
the reading to be plausible and convincing." after their reading. They chose options on a 5-point Likert scale, ranging from 'Strongly Disagree' to 'Strongly Agree' (1 = Strongly Disagree, 2 = Disagree, 3 = Neutral, 4 = Agree, 5 = Strongly Agree)."}
\label{fig:agreeability}
\end{figure}



\subsection{Correlation's in answers}
\label{appendix:correlations_sec}
Figures beginning with Figure \ref{fig:correlations_s1} illustrate correlations between survey participants' responses across different AI safety beliefs, preferred timelines, and familiarity with concepts. Each figure examines how one statement (shown on x-axis) correlates with other statements, preferred AGI development timelines, and AI safety knowledge (1=strongly disagree/never heard of it, 5=strongly agree/know it well). We show the correlation strength with regression lines, and identify statistical significance (p<0.05) when relevant. For example, Figure A1 shows how agreement with ``AGI is too far away to be worth worrying about" relates to other beliefs, revealing key patterns in AI safety skepticism. Refer to Appendix \ref{sec:survey} for the full survey. Refer to Section \ref{design-choices} for the survey design choices. A full list of all (15+) correlation plots like these can be found in our Github.

\begin{minipage}{\textwidth}
\begin{figure}[H]
    \centering
    \includegraphics[width=\textwidth]{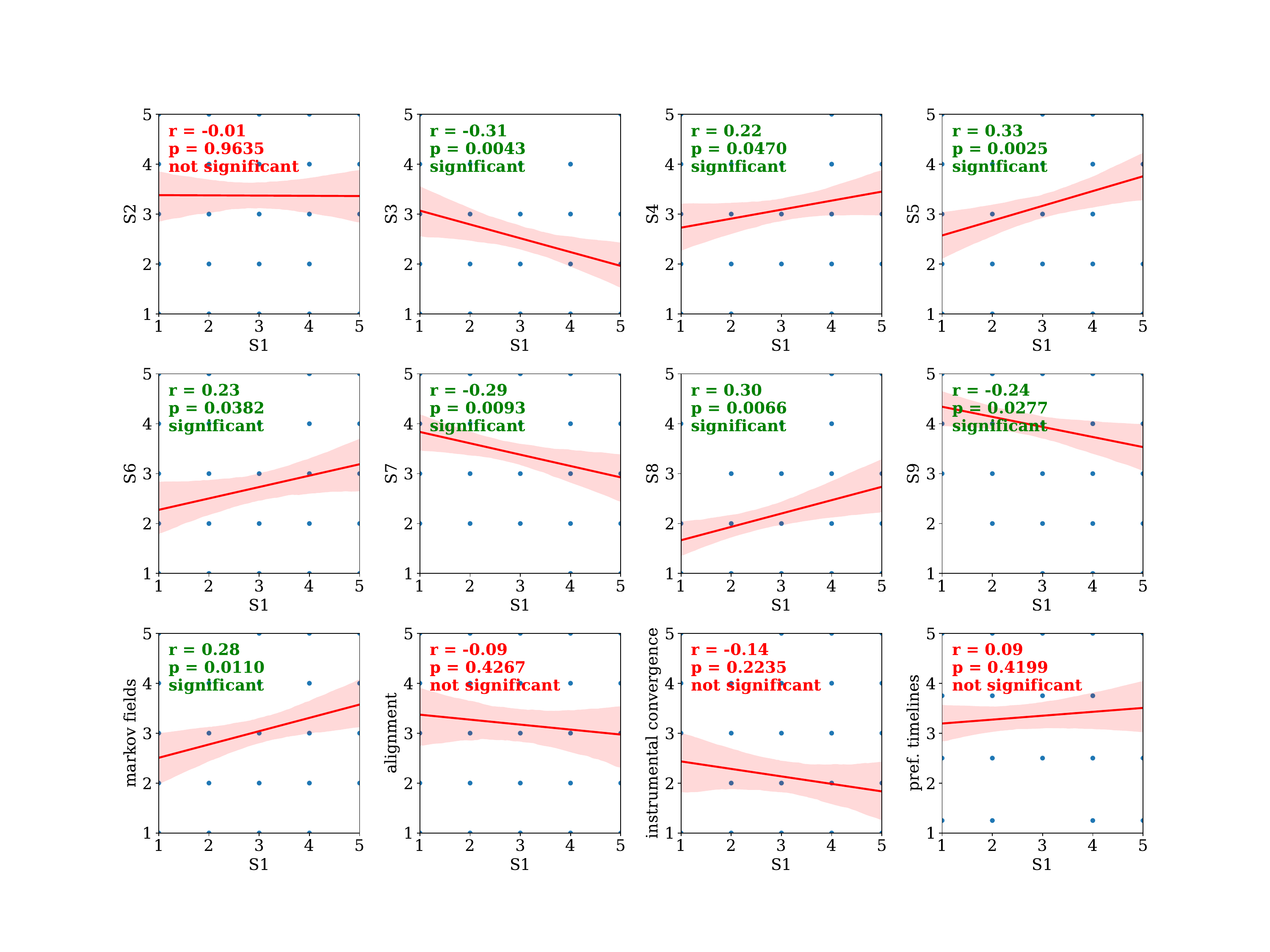}
    \vspace{-20pt} 
    \caption{Correlations with Statement 1: ``AGI is too far away to be worth worrying about"}
    \label{fig:correlations_s1}
\end{figure}

\begin{figure}[H]
    \centering
    \includegraphics[width=\textwidth]{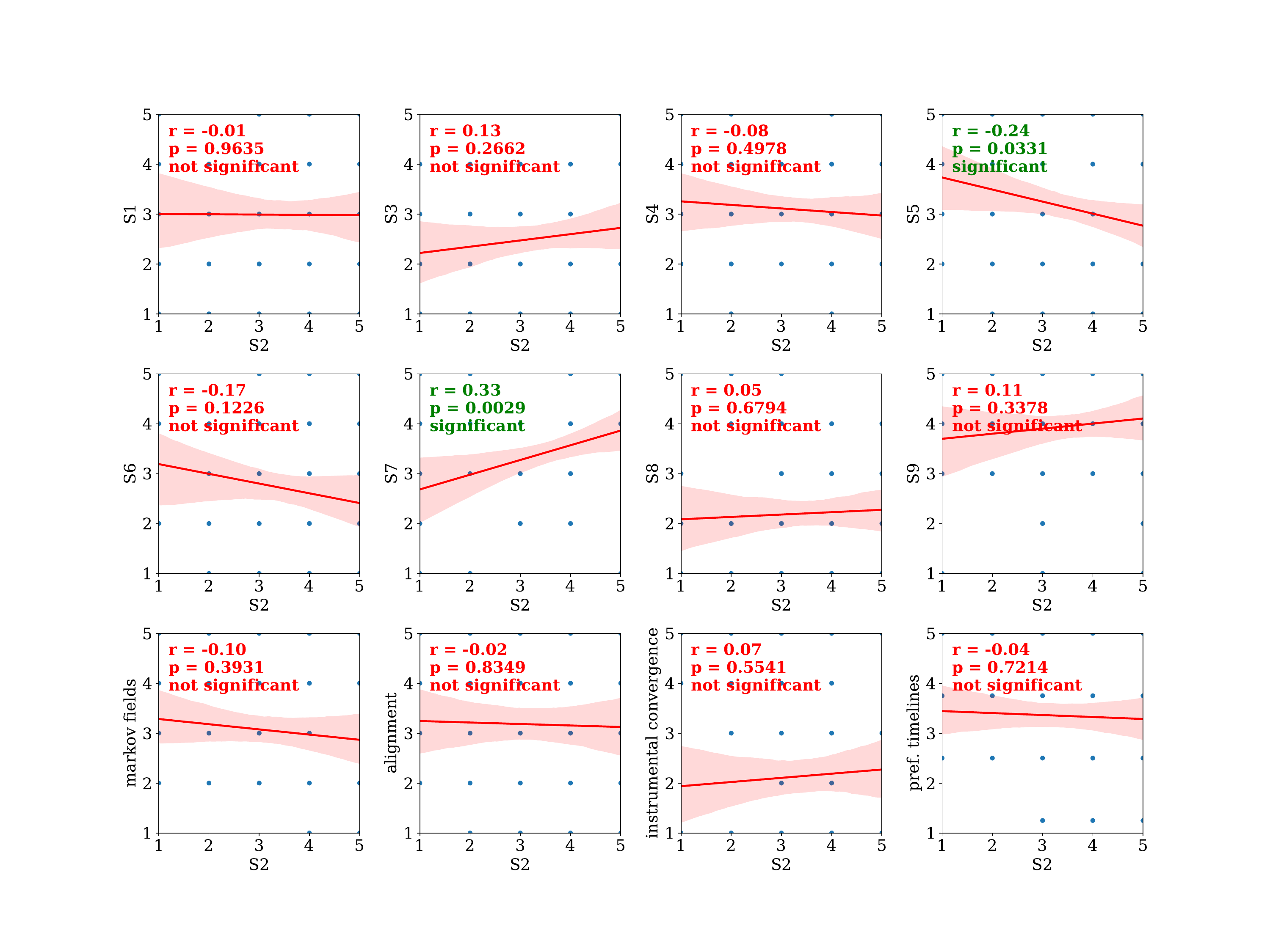}
    \vspace{-20pt} 
    \caption{Correlations with Statement 2: ``Some AI\'s (now or in the future) may be moral patients, with their own welfare that we should care about"}
    \label{fig:correlations_s2}
\end{figure}
\end{minipage}

\end{appendices}


\bibliography{references}

\end{document}